\documentclass{elsart}

\usepackage{amssymb}
\usepackage{graphicx}
\usepackage{placeins}
\usepackage{float}
\usepackage{gensymb}
\usepackage{enumitem}

\begin{document}



\begin{frontmatter}

            
\title{ \rm Production cross sections and angular distributions of neutron-rich rare isotopes 
            from 15 MeV/nucleon Kr-induced collisions: toward the r-process path  }
         

\author{ O. Fasoula$^{a}$, G.A. Souliotis$^{a,*}$, Y.K. Kwon$^{b}$, K. Tshoo$^{b}$}
\author{ M. Veselsky$^{c}$,  S.J. Yennello$^{d}$ and A. Bonasera$^{d,e}$ }

\address{ $^{a}$
           Laboratory of Physical Chemistry, Department of Chemistry,
           National and Kapodistrian University of Athens, Athens, Greece }
              
\address{ $^{b}$  The Rare Isotope Science Project (RISP), Institute for Basic Science, 
                   Daejeon,   Korea }

\address{ $^{c}$ Institute of Experimental and Applied Physics,
                 Czech Technical University, Prague, Czech Republic }

\address{ $^{d}$ Cyclotron Institute, Texas A\&M University,
                     College Station, Texas, USA           }

\address{ $^{e}$ Laboratori Nazionali del Sud, INFN, Catania, Italy }

\address{ $^{*}$ Corresponding author. Email: soulioti@chem.uoa.gr }




\begin{abstract}

We present our recent study of cross sections 
and angular distributions of projectile fragments from heavy-ion reactions at beam energy of 
15 MeV/nucleon.
We studied the production cross sections and the angular distributions of neutron-rich nuclides
from collisions of a $^{86}$Kr (15 MeV/nucleon) beam  with heavy targets ($^{64}$Ni, $^{124}$Sn 
and $^{238}$U).
Experimental data from our previous work at Texas A \& M were compared with model calculations.
Our calculations were based on a two-step approach:  the dynamical stage of the collision was 
described with, first, the phenomenological Deep-Inelastic Transfer model (DIT) and, alternatively,
with the microscopic Constrained Molecular Dynamics model (CoMD). 
The de-excitation of the hot heavy projectile fragments was performed with the Statistical 
Multifragmentation Model (SMM). 
An overall good discription of the available data was obtained with the models employed.
Furthermore, we performed calculations with a radioactive beam of $^{92}$Kr (15 MeV/nucleon) 
interacting with a target of $^{238}$U. We observed that the multinucleon transfer mechanism leads
to extremely  neutron-rich nuclides 
toward and beyond the astrophysical  r-process path. 



\end{abstract}

\end{frontmatter}


\normalsize

\section{Introduction}

The exploration of the nuclear landscape toward the  astrophysical r-process path and the 
neutron drip-line are at the center of interest of the nuclear physics community  
\cite{ndrip0}. Essential to this development  is the efficient production
of neutron-rich nuclides which constitutes a central issue in current and upcoming rare 
isotope beam facilities worldwide
(see, e.g., \cite{FRIB1,FRIB2,GANIL,GSI,RIBF,ARGONNE,EURISOL,RISP,RISP-2013,KOBRA-2016}).

The traditional routes to produce neutron-rich nuclides  are spallation, fission and projectile  
fragmentation \cite{RIB-2013}.
Spallation is an efficient mechanism to produce rare isotopes for ISOL-type techniques
\cite{Spallation}.
Projectile fission is appropriate in the region of asymmetric fission peaks of the light and
heavy fission fragments (see, e.g., \cite{Ufission}).
Finally, projectile fragmentation constitutes a general approach to produce exotic nuclei 
at beam energies above 100 MeV/nucleon (see, e.g., \cite{MSUfrag1,MSUfrag2}).
This approach is, nevertheless, limited by the fact that optimum neutron excess in the fragments 
is achieved by stripping the maximum possible number of protons
(and a minimum possible number of neutrons).
Surpassing the limits of the traditional approaches and reaching out to the neutron dripline 
is highly desirable at present.  Thus, the study of new synthesis routes constitutes a vigorous 
endeavor of the nuclear science community. 

Toward this end, to reach a high neutron-excess in the products, apart from proton stripping,
it is necessary to pickup neutrons from the target.
Such a possibility is offered by  reactions of nucleon exchange  at beam energies
from the Coulomb barrier \cite{Volkov,Corradi} to the Fermi energy
(20--40 MeV/nucleon) \cite{GS-PLB02,GS-PRL03}.
Detailed experimental data in this broad energy range are limited at present 
\cite{Corradi,GS-NIM03,GS-NIM08}.
In multinucleon transfer and deep-inelastic reactions near the Coulomb barrier \cite{Corradi},
the low velocities of the fragments  and the wide angular and ionic charge state distributions
may limit the collection efficiency for the most neutron-rich products.
However, the reactions in the Fermi energy regime (15--35 MeV/nucleon)   
combine  the advantages  of both low-energy (i.e., near and above the Coulomb barrier)  
and high-energy (i.e., above 100 MeV/nucleon)  reactions. 
At this energy, the interaction of the projectile with a neutron-rich target enhances the 
N/Z of the fragments, while the velocities remain  high enough to allow efficient in-flight 
collection and separation.

Our initial experimental studies of projectile fragments from 25 MeV/nucleon reactions 
 of $^{86}$Kr on $^{64}$Ni \cite{GS-PLB02} and $^{124}$Sn \cite{GS-PRL03} indicated 
substantial production of neutron-rich fragments.
Motivated by developments in several facilities that will offer either very intense primary
beams \cite{GANIL,ARGONNE,RISP} at this energy range or re-accelerated rare isotope beams \cite{FRIB2,GANIL,ARGONNE,EURISOL,RISP}, we continued our experimental and theoretical studies 
at 15 MeV/nucleon \cite{GS-PRC11,Fountas-2014,Vonta-2016,Papageorgiou-2018}.

In this contribution, we present our study of the production cross sections and the 
angular distributions of projectile fragments from collisions of 
a $^{86}$Kr (15 MeV/nucleon) beam  with heavy targets.
Data from our experimental work at Texas A\&M are compared  with calculations based on
either the phenomenological deep-inelastic  transfer (DIT) model, or the microscopic constrained
molecular dynamics model (CoMD). 
A overall good description of the experimental data with DIT or CoMD 
was obtained, suggesting the  possibility of using the present theoretical models to predict 
the production of exotic nuclei employing radioactive beams.
As an example, we will show production cross sections and rates of 
neutron-rich isotopes from a radioactive beam of $^{92}$Kr (15 MeV/nucleon)
interacting with a $^{238}$U target.


\section{Outline of Results and Comparisons} 

A detailed presentation of our previously obtained experimental data appear 
in \cite{GS-PRC11} in which the mass spectrometric measurements of production 
cross sections of neutron-rich projectile fragments  from the reactions of a 
15 MeV/nucleon $^{86}$Kr beam with $^{64,58}$Ni and $^{124,112}$Sn targets 
were presented.
We also mention that the experimental data of 25 MeV/nucleon $^{86}$Kr-induced reactions 
are described in detail in our 
articles \cite{GS-PLB02,GS-PRL03,GS-NIM03,GS-NIM08}.

\subsection{Study of Production Cross Sections} 


\begin{figure}[h]                                        
\centering
\includegraphics[width=0.55\textwidth,keepaspectratio=true]{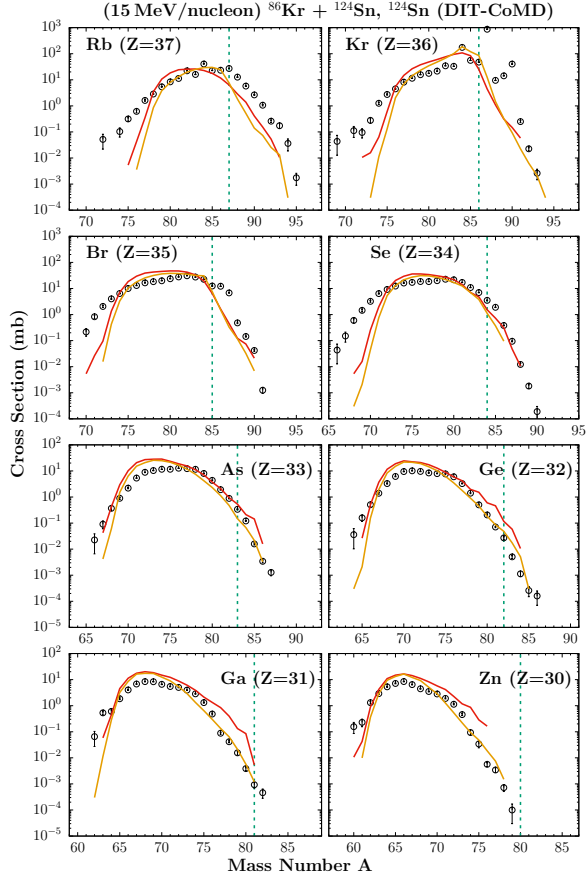}
\caption{ 
Experimental mass distributions (open black circles) of elements with Z = 37--30 
from the reaction $^{86}$Kr(15 MeV/nucleon) + $^{124}$Sn \cite{GS-PRC11} compared with 
DIT/SMM calculations (yellow line) and  CoMD/SMM calculations (red line).
}
\label{fig01_86kr_124sn}
\end{figure}


In Fig. 1, we present the experimental mass distributions of projectile-like fragments
with Z = 37--30 from the reaction $^{86}$Kr (15 MeV/nucleon) + $^{124}$Sn \cite{GS-PRC11}
and compare them with the calculations using the DIT code \cite{DIT,DITm} (solid yellow line)
and the CoMD code \cite{CoMD1,CoMD2} (solid red line). Both dynamical codes were combined with  
the Statistical Multifragmentation Model (SMM) \cite{SMM} used for the  de-excitation of the 
hot quasi-projectiles emerging after the dynamical stage of the reaction.
The results of the calculations are in overall good agreement with the experimental data
especially for isotopes close to the projectile with Z = 36--32.
We note that in some of the Kr isotopes (A=86,89,90), the experimental cross sections 
are especially high because of contamination from elastically scattered beam. 
The overestimation  of the cross sections for n-rich products with Z=31 and 30 from the 
CoMD calculation  may be  related  to issues of low excitation energy,  
as currently calculated in CoMD \cite{Fountas-2014}.

\begin{figure}[t]                                        
\centering
\includegraphics[width=0.55\textwidth,keepaspectratio=true]{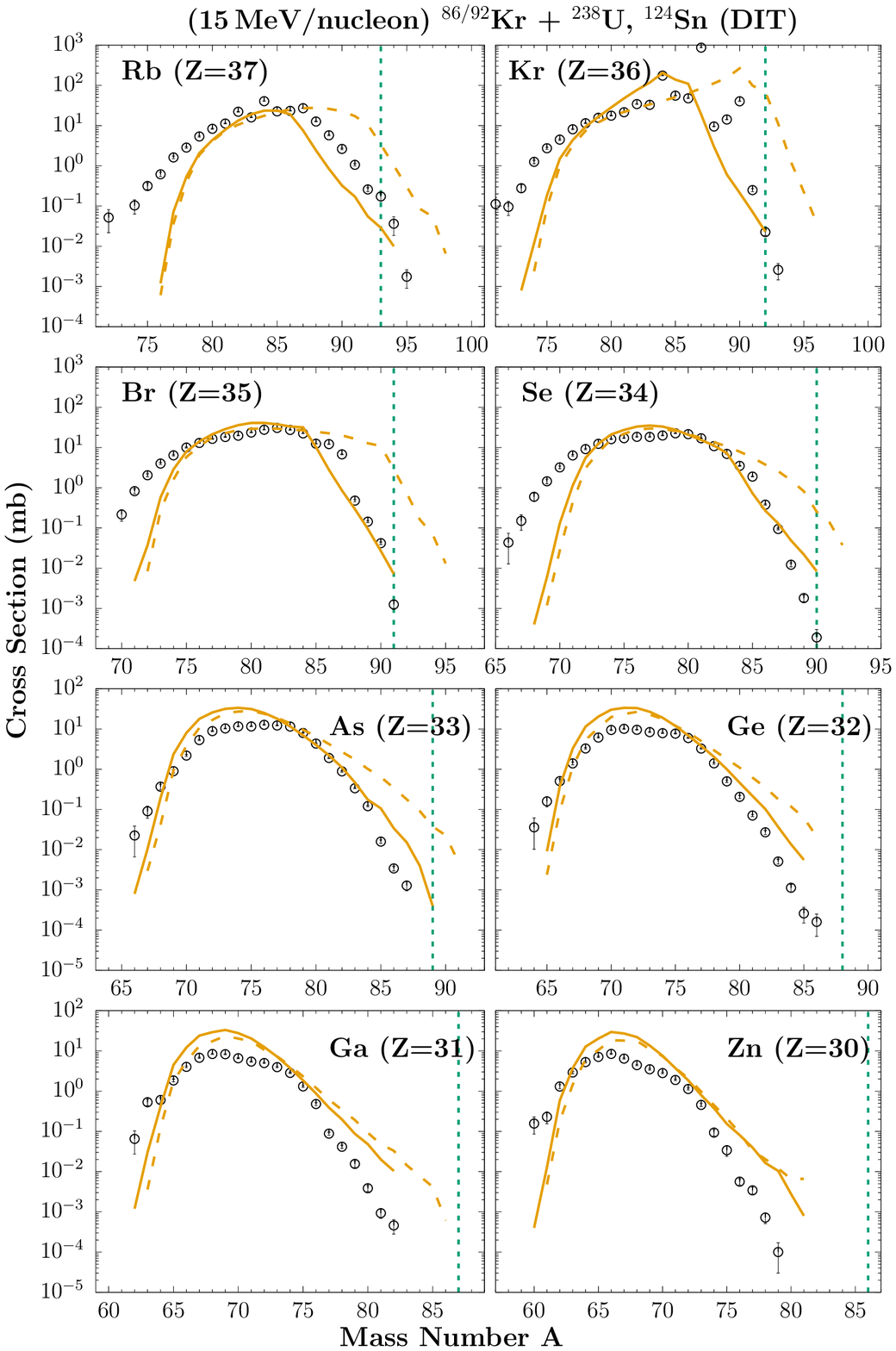}
\caption{ 
DIT/SMM calculated mass distributions of elements with Z = 37--30 
from the reaction of a radioactive beam of $^{92}$Kr (15 MeV/nucleon) interacting with 
a $^{238}$U target (dashed yellow line),  compared to the  DIT/SMM calculations 
of the stable beam  reaction $^{86}$Kr + $^{238}$U (solid yellow line).
The experimental mass distributions (open points) from $^{86}$Kr (15 MeV/nucleon) 
+ $^{124}$Sn \cite{GS-PRC11} are also given for  reference.
}
\label{fig02_86/92kr_238u}
\end{figure}


Furthermore, in Fig. 2 we show the DIT/SMM calculations of projectile-like fragments
from the reaction $^{86}$Kr(15 MeV/nucleon) + $^{238}$U (solid yellow line). 
We have chosen the heaviest and most neutron-rich target available in order to explore 
how far in N/Z we can go with this reaction. 
Also, we chose to study the reaction with the $^{238}$U target, but with a radioactive beam
of $^{92}$Kr(15 MeV/nucleon) (dashed yellow line).
As there are no experimental data with the $^{238}$U target,  we used the data from
$^{86}$Kr (15 MeV/nucleon) + $^{124}$Sn for reference.
We observe that the distributions extend far to the neutron-rich side  especially when 
the radioactive beam is used. This tendency is pronounced for isotopes rather 
close to the projectile (Z=34--36).

We point out that, for, e.g., bromine (Z=35), isotopes that have up to 15 more 
neutrons (A = 96) than the corresponding stable isotope (A = 81) can be obtained. 
Thus,  by using neutron-rich radioactive beams, and via
the mechanism of peripheral multinucleon transfer, we have the possibility
to produce extremely neutron-rich nuclides.
We note that similar observations were made for the reaction with lighter 
projectiles (of Ar and Ca) on the $^{238}$U target in our recent article 
\cite{Papageorgiou-2018}.
We do not have experimental data for the $^{86}$Kr+$^{238}$U reaction, 
but we have plans to study it in the near future. 

\begin{figure}[h]                                        
\centering
\includegraphics[width=0.55\textwidth,keepaspectratio=true]{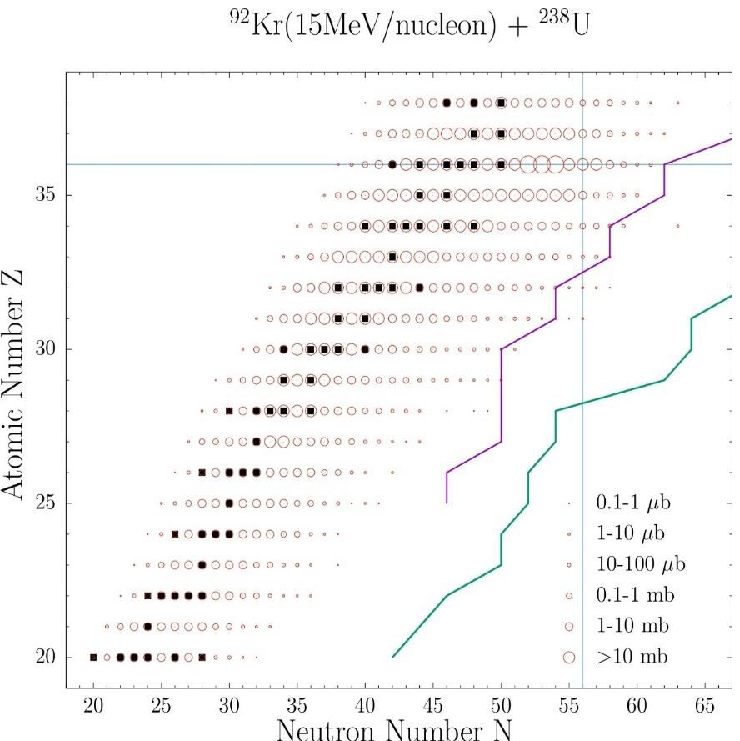}
\caption{ 
   Representation of DIT/SMM calculated production cross sections of projectile fragments from
   the radioactive-beam reaction $^{92}$Kr (15 MeV/nucleon) + $^{238}$U on the Z--N plane.
   Cross section ranges are shown by open circles according to the key.
   Closed squares show the stable isotopes. The purple line shows the
   astrophysical r-process path and the green line shows the neutron drip-line.
   The horizontal and vertical lines  indicate, respectively, the Z and N 
   of the $^{92}$Kr projectile.
}
\label{fig03_92kr_238u_rates}
\end{figure}


A comprehensive presentation of the DIT/SMM calculated production cross sections of 
the projectile fragments from the above reaction on the Z vs N plane
is given in Fig. 3. In this figure, stable isotopes are represented by closed squares, 
whereas fragments  obtained by the $^{92}$Kr+$^{238}$U reaction  are given by the open circles, 
with sizes corresponding to cross-section ranges according to the figure key.
The green line gives the location of the neutron drip-line and the purple line
indicates the expected path of the astrophysical rapid neutron-capture process (r-process).
Using this representation of the previous calculation ,  we clearly observe that the neutron 
pickup products from $^{92}$Kr  reach and even exceed the path of the r-process near Z=30--36.

%

In Table I, we summarize the calculated cross sections and production rates of 
several neutron-rich isotopes from the reaction of the radioactive beam of $^{92}$Kr 
(15 MeV/nucleon) with the $^{238}$U target.
For the rate calculations,  the $^{92}$Kr beam with intensity of 
1.0 pnA (6.2$\times$10$^9$ particles/sec) is assumed to interact with a $^{238}$U target of 
20 mg/cm$^{2}$ thickness. 
We see that it is possible to produce extremely neutron-rich isotopes in these reactions
with the use of re-accelerated radioactive beams, such as $^{92}$Kr,  that will become available 
in  upcoming rare-isotope facilities (e.g. \cite{RISP,RISP-2013,KOBRA-2016}).
Along these lines, we wish to continue this work at the following facilities: a) the Cyclotron Institute of Texas A\&M University using the MARS separator, b) the LNS/Catania using beams from the S800 Cyclotron and employing the MAGNEX spectrometer,  and c) the RISP facility with stable and radioactive beams from the RAON accelerator complex using the KOBRA separator. 


\begin{table}[h]                     
\caption{ Cross sections and rate estimates (last column) of very neutron-rich isotopes
          from the reaction $^{92}$Kr (15 MeV/nucleon) + $^{238}$U.
          For the rates, a radioactive beam of $^{92}$Kr with intensity           
          1.0 particle nA (6.2$\times$10$^9$ particles/s) is
          assumed  to  interact with  a $^{238}$U target  of 20 mg/cm$^{2}$ thickness.
                   }
\vspace{0.5cm}
\begin{center}
\begin{tabular}{cccc}
\hline
\hline
 Rare      & Reaction &  Cross          &  Rate (sec$^{-1}$) \\ 
 Isotope   & Channel  &  Section (mb)   &        \\ \hline
$^{93}$Kr  & -0p+1n &     12               & 3.6$\times$10$^{3}$  \\
$^{94}$Kr  & -0p+2n &    1.3               & 3.9$\times$10$^{2}$  \\
$^{95}$Kr  & -0p+3n &    0.3               &   90   \\
$^{96}$Kr  & -0p+4n &    0.05              &   15   \\

$^{92}$Br  & -1p+1n &    0.8              &   2.4$\times$10$^{2}$  \\
$^{93}$Br  & -1p+2n &   0.2               &   60                   \\
$^{94}$Br  & -1p+3n &   0.07              &   21                   \\
$^{95}$Br  & -1p+4n &   0.02              &    6                   \\
$^{96}$Br  & -1p+5n &   0.008             &    2                   \\

$^{90}$Se  & -2p+0n &     0.25            &   75                  \\
$^{91}$Se  & -2p+1n &     0.14            &   42                  \\
$^{92}$Se  & -2p+2n &     0.05             &  15                  \\
$^{93}$Se  & -2p+3n &     0.02             &   6                  \\

\hline
\hline
\label{RIB-092Kr}
\end{tabular}
\end{center}
\end{table}


\subsection{Study of Angular Distributions} 


After the description of the production cross sections, which have been the main focus 
of our work so far,  we initiated a detailed study of the angular distributions of 
the projectile-like fragments.
In regards to experimental data, we refer again to our previous work \cite{GS-PRC11} 
with the MARS spectrometer.
As described in \cite{GS-PRC11}, the data were obtained at two angle settings,  by
sending the beam on the target at the appropriate angle in the MARS target chamber.
The angle settings were: the 4.0\degree setting and the  7.5\degree setting, 
covering the polar angular ranges of 
$\Delta\theta$=2.2\degree--5.8\degree, and 
$\Delta\theta$=5.6\degree--9.2\degree, respectively.
We remind that approptiate integration of these data provided the total production
cross sections that are reported in \cite{GS-PRC11} and presented in Figs. 1,2.

We start our presentation with the angular distributions of selected projectile fragments 
from the reaction of $^{86}$Kr+$^{124}$Sn at 15 MeV/nucleon (Figures 4,5 and 6).
For this system, the grazing angle is $\theta_{gr}\simeq$ 9.0\degree. (A verical dashed line indicates this angle in the figures.) We, thus, expect that the 7.5\degree angle setting is the appropriate one to efficiently collect the quasi-elastic products, as has been discussed in \cite{GS-PRC11,Fountas-2014}.

The experimental data of the differential cross sections for each nuclide consist of two points:
one at $\theta$=4.0\degree and the other at $\theta$=7.5\degree, each plotted with horizontal error bars indicating the polar angular acceptance of $\Delta\theta$ = 3.6\degree). (The vertical error bars indicate,
as usual, the statistical uncertainties).
The data are compared with calculations employing the DIT model (yellow symbols) and the CoMD model, 
both followed by the SMM model, as also shown in the previous comparisons for the cross sections.
The calculations were binned and presented in angular steps of two degrees. 

Three different of calculations were performed with the CoMD model.
a) The ''standard'' calculation (red symbols) using standard parameters of the CoMD code, as in our 
previous works \cite{Fountas-2014,Papageorgiou-2018}.
b) A calculation with pairing (blue symbols). For this calculation, we introduced a phenomenological 
description of n-n and p-p pairing, according to the recent article \cite{Agodi-2018}. In this description, for two neutrons (or protons ) that have anti-parallel spins and energies near the Fermi energy, an additional attractive interaction was introduced with strength adjusted to reproduce the empirical pairing energy of the Bethe-Weizsacker equation. 
c) a calculation in which the compressibility of cold symmetric nuclear matter was increased to K=272 MeV (purple symbols). We remind that in the "standard" CoMD calculation, the parameters of the effective interaction correspond to a compressibility of K=200 MeV. 

\begin{figure}[t]                                        
\centering
\includegraphics[width=0.55\textwidth,keepaspectratio=true]{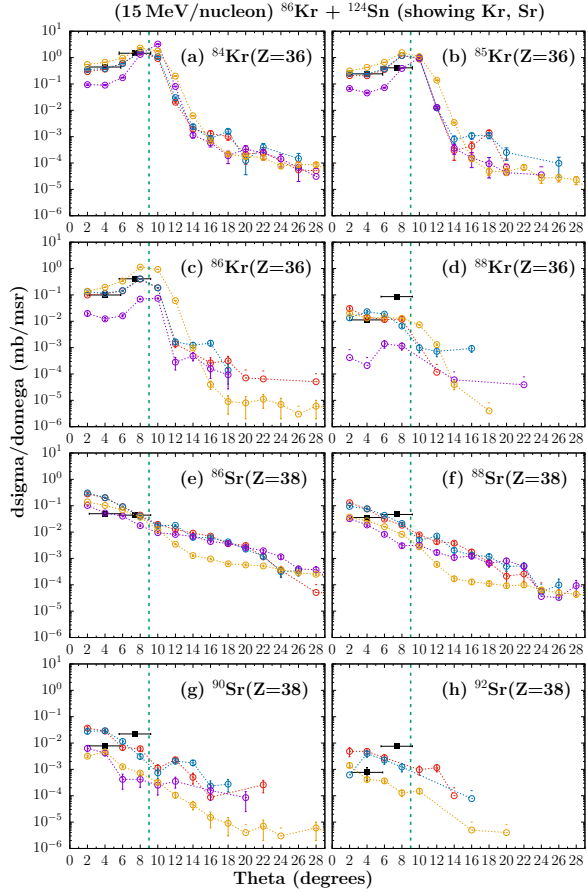}
\caption{ 
Angular distributions of neutron rich isotopes of $_{36}$Kr and $_{38}$Sr 
from the reaction of  $^{86}$Kr (15 MeV/nucleon) with $^{124}$Sn.
Black squares (with horizontal errorbars): experimental data \cite{GS-PRC11}.
Symbols connected with dotted lines: calculations as follows.
Yellow symbols: DIT. Red symbols: standard CoMD. Blue symbols: CoMD with pairing.
Purple symbols: CoMD with compressibility K=272 MeV.
}
\label{fig04_ang_86kr_124sn_z3638}
\end{figure}


In the upper half of Fig. 4, we show the angular distributions for Krypton (Z=36).
For A=86, we see that the data point at $\theta$=7.5\degree is higher than that at 
$\theta$=4.0\degree, as expected.
This step seems to be described by the DIT calculation, despite the larger value of the latter
at $\theta$=8\degree and $\theta$=10\degree. 
The CoMD calculations (in all three variations) appear to describe the shape of the angular distribution 
rather well; the CoMD calculation with K=272 is however lower that the other calculations.
For A=85 and A=84, there is some agreement of the calculations with the data.
However, for A=88 (+2n channel) the calculations cannot describe the step in the data.
This may be related to enhanced neutron-pair transfer, possibly beyond our tentative pairing 
calculation. 

In the lower half of Fig. 4, we show the angular distributions for Strontium (Z=38, +2p channels).
For A=88, we note that the +2p channel does not peak at the grazing angle  as the +2n channel (Fig. 8d).
Moreover, for A=86, the removal of neutrons leads to a flat distribution in the data.
The addition of neutrons, however, leads to angular distributions
peaked at $\theta_{gr}$. This feature cannot be described by DIT or CoMD, and requires further
investigation.

\begin{figure}[H]                                        
\centering
\includegraphics[width=0.55\textwidth,keepaspectratio=true]{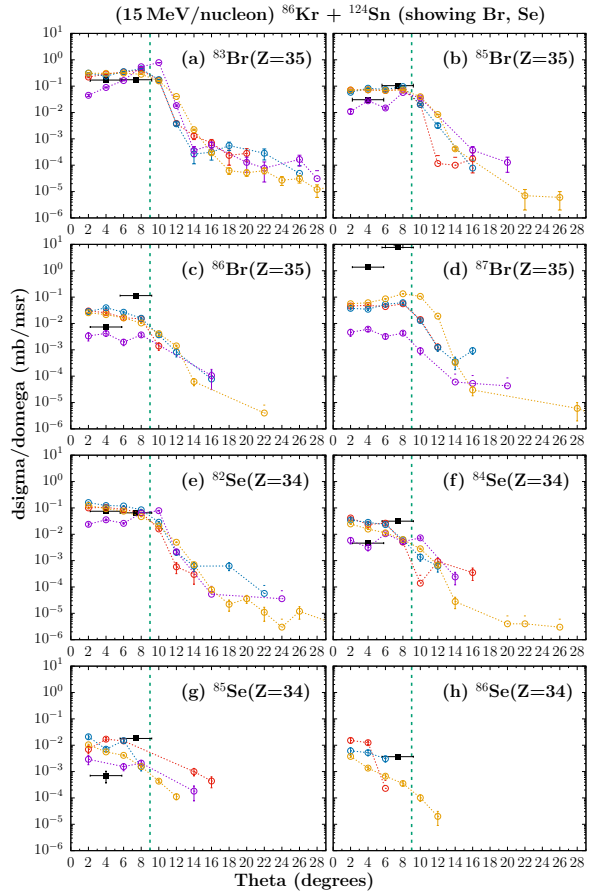}
\caption{ 
Angular distributions of neutron rich isotopes of $_{35}$Br and $_{34}$Se
from the reaction of  $^{86}$Kr (15 MeV/nucleon) with $^{124}$Sn.
Black squares (with horizontal errorbars): experimental data \cite{GS-PRC11}.
Symbols connected with dotted lines: calculations as follows.
Yellow symbols: DIT. Red symbols: standard CoMD. Blue symbols: CoMD with pairing.
Purple symbols: CoMD with compressibility K=272 MeV.
}
\label{fig05_ang_86kr_124sn_z3534}
\end{figure}


In the upper half of Fig. 5,  we show the angular distributions of several isotopes 
of bromine (Z=35).
For A=85 (-1p channel), the calculations describe rather well the $\theta$=7.5\degree data point, 
but are higher than the $\theta$=4.0\degree data point. 
Interestingly, the CoMD calculation with K=272 MeV appears to describe both data points well.
Again, the removal of neutrons (A=83, -1p-2n) results in a nearlyflat angular distribution 
that seems to be described by the calculations.
Concerning the neutron pickup nuclides (A=86, -1p+1n) and (A=87, -1p+2n), the experimental data
are peaked at $\theta_{gr}$ and cannot be described by the DIT and CoMD calculations.
We think that for A=87 (-1p+2n),  the experimental data may be contaminated from background
of  elastically  scattered beam.
Moreover, especially for A=86, we speculate that part of the discrepancy may be due to a 
contribution of a single charge exchange (SCE) process    
that, of course, cannot be described by DIT or CoMD.

In the lower half of Fig. 5, we show the angular distributions of several isotopes of Selenium (Z=34)
(-2p channels). Observations similar to those of the previous figure pertain here.
For A=82 (-2p-2n), the experimental angular distribution is rather flat and is  well described  
by the calculations.
However, for A=84 (-2p),  the experimental angular distribution is again peaked 
at $\theta_{gr}$ and cannot be adequately described by the calculations.
The situation for the angular distributions  of A=85 (-2p+1n) and A=86 (-2p+2n) is similar
to the previous ones.
We may attribute the discrepancy for A=85 in part to a contributing SCE process along 
with a +1n process.
For A=86 (-2p+2n), we speculate that, a DCE  (double charge exchange) process might also 
contribute, along with two successive SCE processes.
The subject of potential contributions of SCE and DCE processes in these heavy-ion reactions
below the Fermi energy is currently under investigation by our group via systematic studies
of the momentum  distributions of the relevant nuclides, and may shed light to interesting 
aspects of the reaction mechanism(s).


In Fig. 6, we show the angular distributions for several isotopes 
of Arsenic (Z=33, -3p channels) in the upper half,  and Germanium (Z=32, -4p channels)
in the lower half.
The angular distributions of these nuclides reveal features analogous to those of the nuclides
shown in the previous figures.
Specifically, the "cold" proton removal channels (-3p and -4p, respectively) are peaked at $\theta_{gr}$ 
(only one point in the data),  whereas the DIT and CoMD calculations show, as before, a monotonic decrease with  increasing angle.
Furthermore, we notice a transition from a $\theta_{gr}$ peaked angular distribution to a flat distribution between A=82 to A=80 for As, and, between A=80 to A=78 for Ge.

\begin{figure}[b]                                        
\centering
\includegraphics[width=0.55\textwidth,keepaspectratio=true]{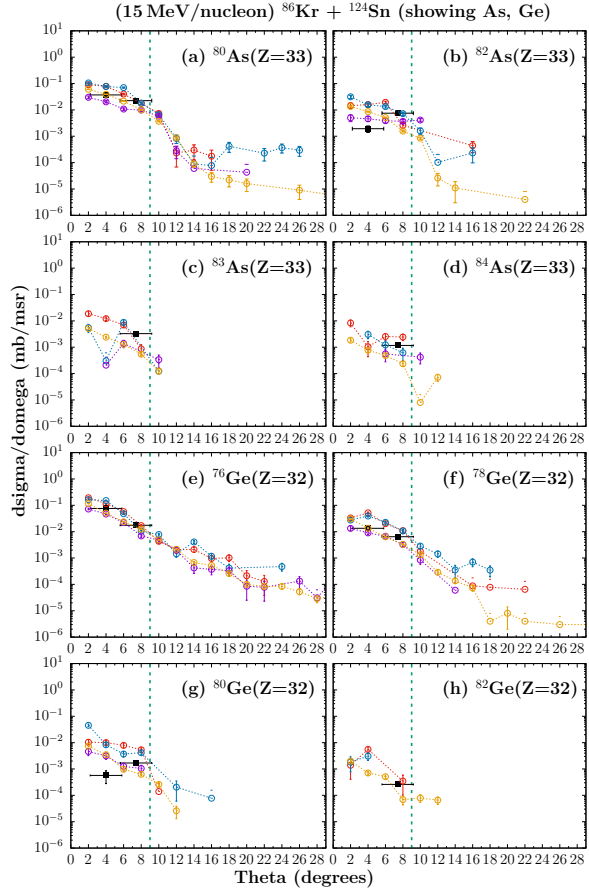}
\caption{ 
Angular distributions of neutron rich isotopes of $_{33}$As and $_{32}$Ge
from the reaction of  $^{86}$Kr (15 MeV/nucleon) with $^{124}$Sn.
Black squares (with horizontal errorbars): experimental data \cite{GS-PRC11}.
Symbols connected with dotted lines: calculations as follows.
Yellow symbols: DIT. Red symbols: standard CoMD. Blue symbols: CoMD with pairing.
Purple symbols: CoMD with compressibility K=272 MeV.
}
\label{fig06_ang_86kr_124sn_z3332}
\end{figure}


We now present the angular distributions of selected projectile fragments from 
the  reaction of $^{86}$Kr (15 MeV/nucleon) + $^{64}$Ni  in Figs. 7 and 8.
For this system, the grazing angle is $\theta_{gr}\simeq$ 6.0\degree. 
(A vertical dashed line indicates this angle in Figs. 7 and 8). 
We expect that the 4.0\degree angular setting is the optimum for the efficient 
collection of quasi-elastic products \cite{GS-PRC11}.
The general feature of the experimental angular distributions is that the 
4.0\degree data points are higher than the corresponding 7.5\degree ones. 
In the upper half of Fig. 7, we show the angular distributions of several isotopes of Krypton (Z=36).
For A=86, the data point at $\theta$=4.0\degree is, as expected,  higher than that at $\theta$=7.5\degree. This step seems to be described by the calculations, which are however lower than the data.
For A=85 and A=84, there is a better agreement of the  calculations with the data. 
However, again in the two-neutron pickup channel, A=88, the calculations are substantially 
lower than the experimental data.
In the lower part of Fig. 7, we present the angular distributions of several isotopes of Strontium 
(Z=38, +2p channels). For these isotopes, the calculations appear to describe the decreasing 
trend of the data with increasing angle. 

\begin{figure}[t]                                        
\centering
\includegraphics[width=0.55\textwidth,keepaspectratio=true]{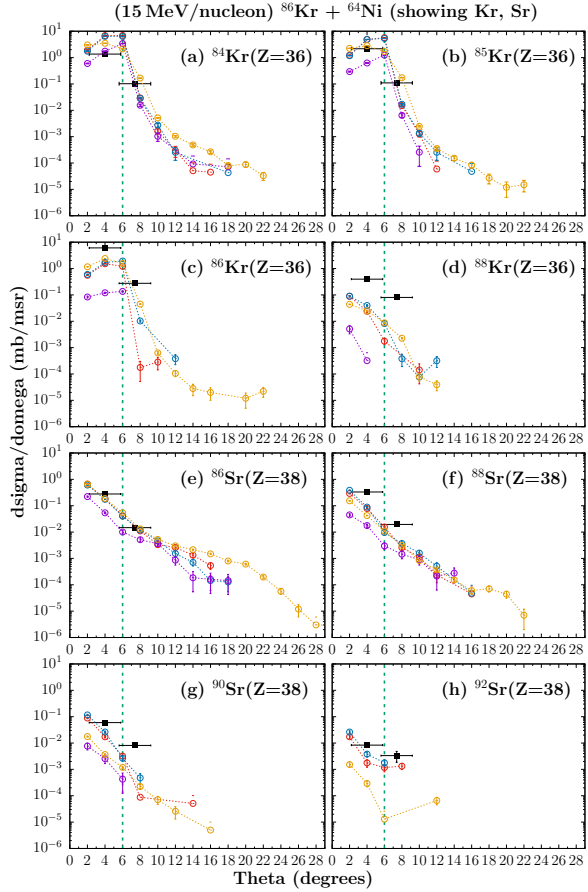}
\caption{ 
Angular distributions of neutron rich isotopes of $_{36}$Kr and $_{38}$Sr
from the reaction of  $^{86}$Kr (15 MeV/nucleon) with $^{64}$Ni.
Black squares (with horizontal errorbars): experimental data \cite{GS-PRC11}.
Symbols connected with dotted lines: calculations as follows.
Yellow symbols: DIT. Red symbols: standard CoMD. Blue symbols: CoMD with pairing.
Purple symbols: CoMD with compressibility K=272 MeV.
}
\label{fig07_ang_86kr_64ni_z3638}
\end{figure}

In the upper part of Fig. 8,  we show the angular distributions of several isotopes 
of Bromine (Z=35, -1p channels). For A=85 (-1p), there is a satisfactory description of the 
data by  the calculations. This is also true for the neutron removal channel, A=83 (-1p-2n ). 
As we move to the neutron pickup channels, A=86 (-1p+1n), and A=87 (-1p+2n), the calculations 
appear to be substantially lower than the experimental data.
In the lower part of Fig. 8, we show the angular distributions of several nuclides of Selenium 
(Z=34, -2p channels).  Observations similar to those of the upper half of figure pertain here.
Specifically, for A=84 (-2p) the calculations appear to follow the data to some extent.
Moving to the neutron removal channel, for A=82 (-2p-2n) the calculations describe the data 
fairly well.
Finally, for the neutron pickup channels, A=85 (-2p+1n), and A=86 (-2p+2n) 
the calculations again follow the trend of the data, but are lower than them.


As a general remark of the angular distribution study,  we point out that
a rather satisfactory overall agreement of the calculations with the  
experimental data was achieved. Several discrepancies were noticed and are 
the subject of further investigation.
As expected, the angular distributions of the Kr+Sn system are more extended
than the Kr+Ni system, which are more forward focused due to inverse kinematics.

Referring to the Kr+Sn system, 
the DIT calculations are rather similar in most cases to the CoMD calculations,
especially to the standard CoMD calculation and the CoMD calculation with the paring.
Specifically, the CoMD calculation with the pairing seems to be slightly higher than
the standard CoMD  calculation for large angles (where, of course, there are no 
experimental data).
Finally, the effect of the compressibility in the CoMD code was investigated by performing 
calculations with K=272 MeV, along with the standard value of K=200 MeV in the other CoMD
calculations.  We observed that with K=272, the CoMD calculated angular distributions are peaked 
more toward the $\theta_{gr}$, as compared to those with the ''standard'' value of K=200 MeV, 
and are thus in better agreement with the data.
However, they have the tendency to be lower that the other CoMD calculations.

\begin{figure}[t]                                        
\centering
\includegraphics[width=0.55\textwidth,keepaspectratio=true]{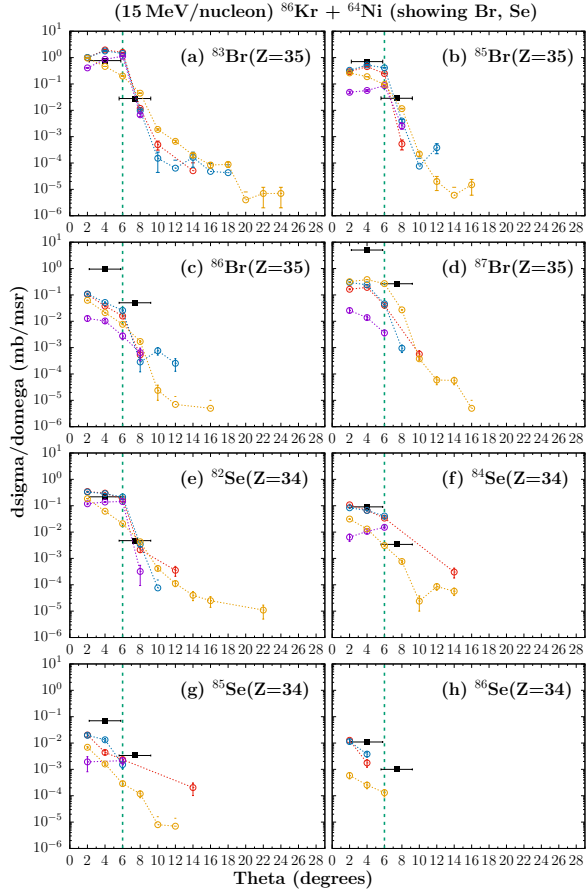}
\caption{ 
Angular distributions of neutron rich isotopes of $_{35}$Br and $_{34}$Se)
from the reaction of  $^{86}$Kr (15 MeV/nucleon) with $^{64}$Ni.
Black squares (with horizontal errorbars): experimental data \cite{GS-PRC11}.
Symbols connected with dotted lines: calculations as follows.
Yellow symbols: DIT. Red symbols: standard CoMD. Blue symbols: CoMD with pairing.
Purple symbols: CoMD with compressibility K=272 MeV.
}
\label{fig08_ang_86kr_64ni_z3534}
\end{figure}


\section{Summary and Conclusions}  


In this contribution, we presented our recent studies of production cross sections 
and angular distributions of projectile fragments from $^{86}$Kr-induced  reactions 
at 15 MeV/nucleon.
We studied the cross sections and the angular distributions of neutron-rich nuclides
from collisions of a $^{86}$Kr (15 MeV/nucleon) beam  with several heavy targets
($^{64}$Ni, $^{124}$Sn  and $^{238}$U).
Experimental data from our previous work at Texas A\&M were compared with model calculations.
The calculations were based on a two-step approach:  the dynamical stage of the collision 
was described with either the Deep-Inelastic Transfer model (DIT), or with the microscopic
Constrained Molecular Dynamics model (CoMD).  The de-excitation of the hot heavy projectile 
fragments was performed with the Statistical Multifragmentation Model (SMM).
We first studied the production cross sections of neutron-rich nuclides from collisions of 
a $^{86}$Kr (15 MeV/nucleon) beam  with $^{124}$Sn and $^{238}$U. 
We also proceeded with calculations with a radioactive beam of $^{92}$Kr (15 MeV/nucleon) 
with $^{238}$U and observed that the multinucleon transfer mechanism leads to very neutron-rich
nuclides in the mass region A=80--100 toward and beyond the 
r-process path. 

Subsequently, we initiated a study of the angular distributions of projectile fragments
from the reactions of $^{86}$Kr (15 MeV/nucleon) with targets of  $^{64}$Ni and $^{124}$Sn.
We compared our experimental data at the two angle settings of the MARS spectrometer, 
namely at 4.0\degree and at 7.5\degree, with detailed calculations using the DIT and CoMD 
models.
Three variants of the CoMD calculations were performed: the standard calculation 
(as in all our previous works),  b) a calculation including a pairing interaction, and c) a
calculation with compressibility K=272 for symmetric nuclear matter.
A rather satisfactory agreement of the calculations with the  experimental data was obtained.
Several discrepancies were noticed and are the subject of further investigation.

Our current studies indicate that heavy-ion reactions below the Fermi energy can be exploited
as an  effective route to access extremely neutron-rich isotopes toward the 
r-process path and the  neutron drip-line.
In this vein, future experiments in several accelerator facilities may be planned that will 
enable a variety of nuclear studies in  unexplored regions of the nuclear chart.   




\end{document}